\documentclass[aps,pra,showpacs,showkeys,twocolumn,superscriptaddress]{revtex4-1}
\usepackage[dvips]{color}
\usepackage{graphicx}
\usepackage{amssymb}
\usepackage{amsmath}
\usepackage[latin1]{inputenc}
\newcommand{\be}{\begin{eqnarray}}
\newcommand{\ee}{\end{eqnarray}}
\begin{document}

\title{Entanglement dynamics via coherent-state propagators}
\author{A. D. Ribeiro}
\author{R. M. Angelo}
\affiliation{Departamento de F\'{\i}sica,Universidade Federal do Paran\'a, 81531-990, Curitiba, PR, Brazil}
\date{\today}

\begin{abstract}
The dynamical generation of entanglement in closed bipartite systems is investigated in the semiclassical regime. We consider a model of two particles, initially prepared in a product of coherent states, evolving in time according to a generic Hamiltonian, and derive a formula for the linear entropy of the reduced density matrix using the semiclassical propagator in the coherent-state representation. The formula is explicitly written in terms of quantities that define the stability of classical trajectories of the underlying classical system. The formalism is then applied to the problem of two nonlinearly coupled harmonic oscillators and the result is shown to be in remarkable agreement with the exact quantum measure of entanglement in the short-time regime. An important byproduct of our approach is a unified semiclassical formula which contemplates both the coherent-state propagator and its complex conjugate.
\end{abstract}


\maketitle

\section{Introduction}\label{int}

Entanglement is one of the most formidable effects of the quantum world. Its puzzling nature, which intrigued the scientific community for a long time, is now being used to accomplish tasks such as quantum information processing, quantum computation, teleportation and quantum cryptography \cite{chuang,horodecki}. Also, its importance has been recognized in the context of several foundational issues underlying the quantum theory, from the explanation of the quantum-classical transition --- and its implications to the measurement problem --- to the understanding of the nonlocal aspects permeating the EPR debate \cite{horodecki,EPR,bell}.

Entanglement is widely believed to be a purely quantum effect with no classical analogue. Despite this common belief, several results have been reported associating the entanglement dynamics with classical quantities. For instance, in Refs.~\cite{kyoko,angelo04,brumer,angelo05} it is shown that the entanglement dynamics can be approximately simulated in the short-time regime by the Liouvillian formalism. In particular, for some specific couplings, the Liouvillian entropy has been shown to reproduce exactly the entropic measure of entanglement for all values of time~\cite{comment0}. In addition, in Ref.~\cite{angelo05} the authors have analytically shown that the short-time dynamics of entanglement does not depend on $\hbar$ for a large class of Hamiltonian systems. Finally, some authors investigated entanglement in the semiclassical regime by means of time-dependent perturbation theory~\cite{gong03,prosen05}. 

The scenario delineated by these works points to a situation in which a statistical theory based on classical trajectories is able to predict the dynamics of a quantity meant to be exclusively quantum \cite{comment1}. This observation leads us to suspect that the entanglement dynamics is initially promoted by mechanisms with well defined classical analogues. Finding out these mechanisms is the main motivation of this paper. We follow, however, a program that is substantially different from the works quoted above, as it is based on semiclassical methods instead of classical statistical theories. Specifically, we propose to derive a semiclassical measure of entanglement in terms of the semiclassical propagator in the coherent-state representation~\cite{scsp1,scsp2,scsp3,aguiar01}.

Recently, a similar calculation has been carried out~\cite{jacquod1,jacquod2} which differs from ours in some important aspects. First, the approach adopted there was based on the Van-Vleck semiclassical propagator~\cite{vv}, 
\begin{equation}
K_{\mathrm{vv}}(q_2,q_1,T) = 
\sum_{\mathrm{traj.}} 
\left|\frac{1}{2\pi \hbar}
\frac{\partial^2 S(q_2,q_1,T)}{\partial q_2\partial q_1} 
\right|^{\frac12}
e^{ \frac{i}{\hbar}S(q_2,q_1,T) }.\qquad
\nonumber
\end{equation}
This is a semiclassical formula for the one-dimensional quantum propagator $\langle q_2| e^{-i\hat{H}T/\hbar} |q_1 \rangle$ in the position representation. This formula depends only on the classical trajectories of an underlying classical dynamics connecting the initial position $q_1$ to the final position $q_2$, during the time interval $T$. The function $S(q_2,q_1,T)$ is the classical action of the trajectory and the sum runs over all trajectories satisfying the boundary conditions. Our approach, on the other hand, is based on a semiclassical propagator formulated in the coherent-state representation, which has the advantage of offering a straightforward extension to systems with spin degrees of freedom~\cite{spin1,spin2,spin3,spin4,jpaspin}.

A second important difference relies on the fact that our approach does not employ any averaging over the initial conditions in phase space. Even though in Refs.~\cite{jacquod1,jacquod2} this statistical procedure is claimed to be nonrestrictive we believe it is not mandatory from a physical point of view. The only approximations used here are those usually associated to the method of the stationary phase. 

Finally, we observe that, contrary to the Van-Vleck propagator, the coherent-state propagator is generally determined by  {\em complex} trajectories and actions. This introduces an additional technical difficulty, namely, that the complex conjugate of the semiclassical propagator does not have a straightforward interpretation. Actually, this turns out to be an interesting mathematical issue to be understood in the context of general applications of the coherent-state propagator. In this paper we formulate and address this problem as a preliminary step towards the derivation of a semiclassical formula for the entanglement.

This article is organized as follows. In Sec.~\ref{prop_xi}, we present the main ingredients of the original formula of the semiclassical propagator in coherent states and then extend it to contemplate also the complex conjugate of the propagator in a unified formalism. We then proceed to calculate the semiclassical purity in Sec.~\ref{semiP}, where a further extension of the semiclassical formula of the propagator is required to accommodate bipartite systems. In Sec.~\ref{example} we present a study of case; the formalism is applied to the problem of two nonlinearly interacting oscillators. Finally, in Sec.~\ref{final} we summarize and conclude the work.

\section{Semiclassical propagator in the coherent-state representation \label{prop_xi}}

The aim of this section is twofold. Firstly, we briefly review some of the main aspects of the semiclassical formula of the coherent-state propagator. For a subsidiary literature on this representation we refer to Refs.~\cite{klauderb,perelomov,gilmore}. Secondly, we show how to extend the formalism so as to semiclassically approach both the propagator and its complex conjugate in a unified mathematical structure. In this sense, our approach intends to offer a generalization of the formula derived in Ref.~\cite{aguiar01}.

We start, as a motivating question, with the general problem of calculating the expectation value of an arbitrary operator $\hat A$ via semiclassical propagator. If the initial state of the system is the coherent state $|z_0\rangle$, then at the instant $T$ the mean value $\langle \hat A\rangle_T=\langle z_0|e^{i\hat H T/\hbar}\,\hat A\, e^{-i\hat H T/\hbar}|z_0\rangle$ can be written in terms of the propagator as
\begin{eqnarray}
\langle \hat A \rangle_T=\int \frac{d^2z_2}{\pi}\frac{d^2z_1}{\pi}K(z_0,z_1,-T)\,A(z_1,z_2)\, K(z_2,z_0,T),
\nonumber
\end{eqnarray} 
where $A(z_1,z_2)=\langle z_1 |\hat A|z_2\rangle$ and
\be
K(z_0,z_1,-T)=\langle z_0| e^{i\hat H T/\hbar} |z_1\rangle=K^*(z_1,z_0,T).\nonumber
\label{K*}
\ee
The complex conjugate of the propagator $K^*$ is going to be present whenever measurable quantities are regarded. However, to the best of our knowledge, there is no prescription on how to obtain the semiclassical version of this object in the coherent-state representation. But should not we simply take the complex conjugate of the semiclassical propagator? We opt here by a more careful strategy that preserves both the interpretation of the critical trajectories and the rigor of the original derivation in Ref.~\cite{aguiar01}. 

\subsection{The coherent-state propagator \label{K_aguiar}}

In Ref.~\cite{aguiar01} it is shown that the semiclassical formula of the coherent-state propagator,
\begin{equation}
K(z_2,z_1,T)= 
\langle z_2|e^{-i\hat HT/\hbar} |z_1\rangle,
\end{equation}
depends only on {\em complex} trajectories of an auxiliary classical system governed by the Hamiltonian function $H(v,u)$, which is to be built according to the prescription 
\be
H(v,u)=\left[\langle z| \hat H|z\rangle\right]_
{\scriptsize \begin{array}{l} z  \,\,\,\to  u \\ z^*  \to  v \end{array}}.
\label{H}
\ee
That is, to find $H(v,u)$, one evaluates $\langle z|\hat H|z\rangle$ and replace $z$ and $z^*$ by $u$ and $v$, respectively. The usual classical variables $q$ and $p$ are related to the variables $u$ and $v$ through
\begin{equation}
u = \frac{1}{\sqrt2} \left( \frac{q}{b} +\frac{ip}{c}\right)
\quad \mathrm{and} \quad
v = \frac{1}{\sqrt2} \left( \frac{q}{b} -\frac{ip}{c}\right),
\label{nv}
\end{equation}
where $b$ and $c$, satisfying $b\,c=\hbar$, are related to the variances of the coherent state along the position and momentum axes. Hamilton's equations written in terms of $u$ and $v$ become
\begin{equation}
\dot u = -\frac{i}{\hbar} \frac{\partial H}{\partial v}
\quad \mathrm{and} \quad
\dot v = \frac{i}{\hbar} \frac{\partial H}{\partial u}.
\label{Keqm}
\end{equation}
Trajectories contributing to the semiclassical propagator must satisfy the boundary conditions 
\begin{equation}
u(0)=z_1
\quad\mathrm{and}\quad
v(T)=z_2^*.
\label{bc1d} 
\end{equation}

A careful inspection of the dynamical structure defined by Eqs.~\eqref{nv}-\eqref{bc1d} reveals why the classical variables $q$ and $p$ must be complex. Since the boundary conditions given by Eq.~\eqref{bc1d} and the evolution time $T$ are both fixed from the outset, it is not possible to find {\em in general} a classical trajectory satisfying that many conditions simultaneously, unless $q$ and $p$ are allowed to be complex numbers. This is the motivation for the change of variables $(z^*,z)\to (v,u)$.

Having found the proper trajectory we can evaluate its complex action,
\begin{equation}
\mathcal{S}(z_2^*,z_1,T) = 
\int_0^T 
\left[\frac{i\hbar}{2} 
\left( \dot u v - u \dot v \right) 
- H(v,u)
\right]dt 
-\Lambda,
\label{S}
\end{equation}
where $\Lambda = \frac{i\hbar}{2}\left[u(0)v(0)+u(T)v(T)\right]$,
and the function 
\begin{equation}
\mathcal{G}(z_2^*,z_1,T) = 
\frac12 \int_0^T 
\left( \frac{\partial^2 H(v,u)}{\partial u \partial v} \right)dt. 
\label{G}
\end{equation}
The semiclassical propagator is then given by
\begin{equation}
\mathcal{K}(z_2^*,z_1,T) = \mathcal N \sum_{\mathrm{traj.}} 
\left( \frac{i}{\hbar} 
\frac{\partial^2\mathcal{S}}{\partial z_2^* \partial z_1}
\right)^{1/2} 
e^{\frac{i}{\hbar} \left(\mathcal S + \mathcal G \right)},
\label{K_original}
\end{equation}
where $\mathcal N = \exp(-\frac{1}{2} |z_2|^2 -\frac{1}{2} |z_1|^2)$. Some comments about Eq.~(\ref{K_original}) are in order. First, it is worth mentioning that $\mathcal{K}$ is obtained through a quadratic approximation around critical paths --- the complex classical trajectories --- of $K$, expressed in the path integral formalism. Second, it is explicitly indicated that, in principle, one should sum contributions of all trajectories satisfying the boundary conditions. Third, the label $z_2$ of $K$ is written as $z_2^*$ in $\mathcal{K}$ as the trajectories depend only on the value of $z_2^*$ instead of $z_2$. In the right hand side of Eq.~(\ref{K_original}), the only dependence on $z_2$ lies in~$\mathcal N$.

The difficulties to get a semiclassical expression for $\mathcal{K}^*$ directly from Eq.~(\ref{K_original}) can be better appreciated at this point. Contrary to the Van-Vleck propagator, the functions $\mathcal S$ and $\mathcal G$ and the classical variables $u$ and $v$ are all complex. Then, taking the complex conjugate of Eq.~(\ref{K_original}) implies working with the complex conjugate of these functions, which although well defined mathematically may not offer a straightforward interpretation from the point of view of the quantum-classical connection.

Next we address this issue preserving the mathematical structure that was carefully derived and extensively discussed in Ref.~\cite{aguiar01}. 

\subsection{Unified semiclassical formula \label{unified}}

Let us consider now the generic propagator
\be
K_{\xi}(z_2,z_1,T)=\langle z_2|e^{-i \xi \hat H T/\hbar}|z_1\rangle,
\label{K_xi}
\ee
where $\hat H$ is a time-independent Hamiltonian, $\xi=\pm 1$, and the kets $|z_1\rangle$ and $|z_2\rangle$ are coherent states. Clearly, $K_-(z_2,z_1,T)=K_+^*(z_1,z_2,T)$. In this sense, Eq.~\eqref{K_xi} contemplates propagators and their complex conjugates in a unified formula.  In addition, we see that $K_{\xi}(z_2,z_1,T)$ can be obtained from $K(z_2,z_1,T)$ by means of the change $\hat{H}\to\xi\hat{H}$~\cite{comment2}. Furthermore, since $\xi$ is nothing but a real constant, the mathematical structure previously delineated readily applies, provided that we consistently employ the mentioned change. 

We start our program of implementing the change $\hat{H}\to \xi\hat{H}$ with Hamilton's equations \eqref{Keqm}. We get
\be
\dot u = -\frac{i\xi}{\hbar} \frac{\partial H}{\partial v}
\quad \mathrm{and} \quad
\dot v = \frac{i\xi}{\hbar} \frac{\partial H}{\partial u}.
\label{Keqmxi}
\ee
This changes the interpretation of $u$ and $v$, making their roles swap in the dynamics depending on the value of $\xi$. In order to avoid this issue we define the generalized time
\be
t_{\xi}\equiv \xi\,t+(1-\xi)T/2.
\label{t_xi}
\ee
Explicitly, we see that $t_+=t$, but for $\xi=-1$ we get $t_-=T-t$. This strategy allows us to preserve the equations of motion in the same form as Eqs.~\eqref{Keqm},
\be
\frac{d u_{\xi}}{dt_\xi} = -\frac{i}{\hbar} \frac{\partial H}{\partial v_{\xi}}
\quad \mathrm{and} \quad
\frac{d v_{\xi}}{dt_\xi} = \frac{i}{\hbar} \frac{\partial H}{\partial u_{\xi}},
\label{eqm_xi}
\ee
where $u_{\xi}$ and $v_{\xi}$ are defined by
\be
u_{\xi}(t_\xi) 
\equiv u(t(t_\xi))
\quad \mathrm{and} \quad
v_{\xi}(t_\xi) 
\equiv v(t(t_\xi)), 
\label{uv_xi}
\ee
with $t(t_\xi)$ given by the inverse of Eq.~\eqref{t_xi}, and
\be
H(v_{\xi},u_{\xi})=
\left[\langle z| \hat H|z\rangle\right]
_{\scriptsize \begin{array}{l} z \,\,\, \to  u_{\xi} \\ z^*  \to  v_{\xi} \end{array}}.
\label{H_xi}
\ee
In terms of the new functions, the boundary conditions given by Eq.~\eqref{bc1d} read
\be
u_{\xi}\left((1-\xi)\textrm{\scriptsize{$\frac{T}{2}$}}\right)=z_1
\quad\mathrm{and}\quad
v_{\xi}\left((1+\xi)\textrm{\scriptsize{$\frac{T}{2}$}}\right)=z_2^*,\quad
\label{bc} 
\ee
or, equivalently,
\be
\begin{array}{ccc}
u_+(0)=z_1, & \quad & v_+(T)=z_2^*, \\ \\
u_-(T)=z_1, & \quad & v_-(0)=z_2^*. 
\end{array}
\label{bc_exp}
\ee

We focus now on the functions $\mathcal{S}_{\xi}(z_2^*,z_1,T)$ and $\mathcal{G}_{\xi}(z_2^*,z_1,T)$, the extended forms of Eqs.~\eqref{S} and~\eqref{G}. Relations~\eqref{uv_xi} give us the rule to rewrite $\mathcal{S}_{\xi}$ and $\mathcal{G}_{\xi}$ in terms of $u_{\xi}(t_{\xi})$ and $v_{\xi}(t_{\xi})$. We then change the variable of integration from $t$ to $t_{\xi}$ and the limits of integration to $(1-\xi)T/2$ and $(1+\xi)T/2$. Finally, we replace the dummy variable $t_{\xi}$ by $t$ and use the identity
\be
\int\limits_{(1-\xi)\frac{T}{2}}^{(1+\xi)\frac{T}{2}} F(t)\, dt=\xi\int\limits_{0}^{T} F(t)\, dt,\nonumber 
\ee
which holds for any $F(t)$ as far as $\xi=\pm 1$. This procedure allows us to write
\begin{eqnarray}
\mathcal{G}_{\xi} = 
\frac{\xi}{2} \int_{0}^{T} \left( \frac{\partial^2 H(v_{\xi},u_{\xi})}{\partial u_{\xi} \,\partial v_{\xi}} \right)dt
\label{Gcal}
\end{eqnarray}
and
\begin{eqnarray}
\mathcal{S}_{\xi} =\xi
\int_{0}^{T} 
\left[\frac{i\hbar}{2} 
\left( \dot u_{\xi} v_{\xi} - u_{\xi} \dot v_{\xi} \right) 
- H(v_{\xi},u_{\xi})
\right]dt 
-\Lambda_\xi,\qquad
\label{Scal}
\end{eqnarray}
where $\Lambda_\xi = \frac{i\hbar}{2}\left[u_{\xi}''\,v_{\xi}''+u_{\xi}'\,v_{\xi}'\right]$. For the sake of compactness of the notation we have introduced the following (double) primed variables:
\be
\begin{array}{lll}
u_{\xi}'\equiv u_{\xi}(0), & \quad & u_{\xi}''\equiv u_{\xi}(T), \\ \\
v_{\xi}'\equiv v_{\xi}(0), & \quad & v_{\xi}''\equiv v_{\xi}(T).
\end{array}
\label{uv'}
\ee
The notation is such that prime (double prime) refers always to the initial (final) instant. Notice by \eqref{uv'} and~\eqref{bc} that, while $u_+'=u_-''=z_1$ and $v_+''=v_-'=z_2^*$, the variables $u_+''$, $u_-'$, $v_+'$, and $v_-''$ are not fixed by the boundary conditions~\eqref{bc}. They are obtained once the solution for the trajectory has been found.

The {\em semiclassical propagator} in the coherent-state representation is then finally written as
\begin{eqnarray}
\mathcal{K}_{\xi}(z_2^*,z_1,T) = \mathcal N \sum_{\mathrm{traj.}} 
\left( \frac{i}{\hbar} 
\frac{\partial^2\mathcal{S}_{\xi}}{\partial z_2^* \partial z_1}
\right)^{1/2} 
e^{\frac{i}{\hbar} \left(\mathcal S_{\xi} + \mathcal G_{\xi} \right)},\quad
\label{Ksemi}
\end{eqnarray}
where $\mathcal N = \exp(-\frac12 |z_2|^2 -\frac12 |z_1|^2)$ remains unchanged. It is worth noticing that for $\xi=+1$ the original formalism is fully reproduced.

Finally, concerning the complex action $\mathcal{S}_{\xi}(z_2^*,z_1,T)$ it satisfies the following useful relations:
\be
\begin{array}{lcl}
\displaystyle
u_{+}''= \frac{i}{\hbar} 
\frac{\partial \mathcal S_{+}}{\partial z_2^*}=\frac{i}{\hbar} 
\frac{\partial \mathcal S_{+}}{\partial v_+''},
&&
\displaystyle
v_{+}'= \frac{i}{\hbar} 
\frac{\partial \mathcal S_{+}}{\partial z_1}=\frac{i}{\hbar} 
\frac{\partial \mathcal S_{+}}{\partial u_{+}'},\\\\
\displaystyle
u_{-}'= \frac{i}{\hbar} 
\frac{\partial \mathcal S_{-}}{\partial z_2^*}=\frac{i}{\hbar} 
\frac{\partial \mathcal S_{-}}{\partial v_-'},
&&
\displaystyle
v_{-}''= \frac{i}{\hbar} 
\frac{\partial \mathcal S_{-}}{\partial z_1}=\frac{i}{\hbar} 
\frac{\partial \mathcal S_{+}}{\partial u_{-}''},
\end{array}
\label{ds}
\ee
and
\begin{eqnarray}
\frac{\partial \mathcal S_{\xi}}{\partial T} = -\xi H(v_{\xi}',u_{\xi}') = -\xi H(v_{\xi}'',u_{\xi}'').
\end{eqnarray}
In addtion, using Eqs.~\eqref{ds}, the prefactor of $\mathcal{K}_{\xi}(z_2^*,z_1,T)$ can be written as a function of the elements of the tangent matrix~$\mathrm{M}_{\xi}$ defined by
\begin{eqnarray}
\left(\begin{array}{l}
\delta u_{\xi}'' \\ \delta v_{\xi}''
\end{array}\right) =
\mathrm{M}_{\xi} 
\left(\begin{array}{l}
\delta u_{\xi}' \\ \delta v_{\xi}'
\end{array}\right)=
\left(\begin{array}{cc}
{M}_{uu}^{\xi} & {M}_{uv}^{\xi} \\
{M}_{vu}^{\xi} & {M}_{vv}^{\xi} 
\end{array}\right) 
\left(\begin{array}{l}
\delta u_{\xi}' \\ \delta v_{\xi}'
\end{array}\right).\qquad
\label{monodK}
\end{eqnarray}
One can show that
\begin{equation}
\frac{i}{\hbar}
\frac{\partial^2 {S}_+}{\partial z_2^*  \partial z_1} =
\frac{1}{M_{vv}^{+}}
\quad\mathrm{and}\quad
\frac{i}{\hbar}
\frac{\partial^2 {S}_-}{\partial z_2^*  \partial z_1} =
\frac{1}{M_{uu}^{-}}, 
\label{prefactor}
\end{equation}
and
\begin{eqnarray}
\begin{array}{rl}
{M}_{uu}^{+} =&
\displaystyle
\frac{i}{\hbar}\left[
\frac{\partial^2\mathcal{S}_{+}}{\partial u_{+}' \partial v_{+}''} - 
\frac{\partial^2\mathcal{S}_{+}}{\partial{v_{+}''}^2} 
\left( 
\frac{\partial^2\mathcal{S}_{+}}{\partial v_{+}'' \partial u_{+}'} 
\right)^{-1}
\frac{\partial^2\mathcal{S}_{+}}{\partial {u_{+}'}^2} 
\right],
\\\\
{M}_{uv}^{+} =&
\displaystyle
\frac{\partial^2\mathcal{S}_{+}}{\partial{v_{+}''}^2} 
\left( 
\frac{\partial^2\mathcal{S}_{+}}{\partial v_{+}'' \partial u_{+}'} 
\right)^{-1},
\\\\
{M}_{vu}^{+} =&
\displaystyle
-\left( 
\frac{\partial^2\mathcal{S}_{+}}{\partial v_{+}'' \partial u_{+}'} 
\right)^{-1}
\frac{\partial^2\mathcal{S}_{+}}{\partial{u_{+}'}^2}.
\end{array}
\label{M+}
\end{eqnarray}
Another set of three equations relating derivatives of $\mathcal{S}_{-}$ with elements of $\mathrm{M}_{-}$ can be obtained by simultaneously replacing $+$, $u$ and $v$ in Eqs.~\eqref{M+} with $-$, $v$ and $u$, respectively. The reason to write the prefactor in terms of elements of the tangent matrix is the ease of handling them in several situations, specially in numerical treatments.

The classical structure we have proposed is such that a trajectory $(u_{\xi}(t_{\xi}),v_{\xi}(t_{\xi}))$  is solution of the equations of motion in terms of a proper time scale $t_{\xi}$. The interpretation of a forward time evolution from $0$ to $T$ is preserved, but while $\mathcal{K}_+(z_2^*,z_1,T)$ depends on a trajectory that propagates from $z_1$ to $z_2^*$, $\mathcal{K}_-(z_2^*,z_1,T)$ depends on one propagating from $z_2^*$ to $z_1$. In this sense, comparing with the case in which $\xi=+1$, trajectories for $\xi=-1$ can also be interpreted in terms of a backward time evolution, which is compatible with the intuition one may construct from the exact relation $K_-(z_2,z_1,T)=K_+(z_2,z_1,-T)$.

The set of equations given in this section defines the general recipe to obtain the semiclassical version $\mathcal{K}_{\xi}$ of the exact propagator $K_{\xi}$ given by Eq.~\eqref{K_xi}. As such, this unified formalism constitutes the first important contribution of this work. All the formal details involved in the derivation of original formulas, specially those associated to the stationary phase method, can be found in Ref.~\cite{aguiar01} for the case in which $\xi=+1$.

\subsection*{A simple example: harmonic oscillator}

In order to clarify the notation and illustrate the adequacy of the formalism, we calculate the semiclassical version of $K_{\xi}(z_2,z_1,T)$ for the harmonic oscillator Hamiltonian, $\hat H_{\rm ho}$. According to Eq.~\eqref{H_xi} the classical Hamiltonian results
\be
H_{\rm ho}(v_{\xi},u_{\xi})= \hbar \omega\left(v_{\xi} u_{\xi}+\frac{1}{2}\right),
\nonumber
\ee
where we have adopted as the coherent-state basis exactly that one associated to $\hat H_{\rm ho}$. Then, from Eq.~(\ref{eqm_xi}) one gets $\dot{v}_{\xi}=i\omega v_{\xi}$ and $\dot{u}_{\xi}=- i\omega u_{\xi}$, whose solutions read 
\be
v_{\xi}(t_\xi)=C_v^{\xi}e^{i\omega t_\xi}
\quad\mathrm{and}\quad 
u_{\xi}(t_\xi)=C_u^{\xi}e^{-i\omega t_\xi}.
\nonumber
\ee
From Eqs.~\eqref{bc} and~\eqref{t_xi} we get
\be
C_v^{\xi}=z_2^*e^{-i\omega(1+\xi)T/2}
\quad\mathrm{and}\quad
C_u^{\xi}=z_1e^{i\omega (1-\xi)T/2},
\nonumber
\ee
so that
\be
\begin{array}{l}
v_{\xi}(t_\xi)=z_2^*e^{i\omega [t_\xi-(1+\xi)T/2]}, \\ \\
u_{\xi}(t_\xi)=z_1 e^{-i\omega [t_\xi-(1-\xi)T/2]}.
\end{array}\nonumber
\label{solutions}
\ee
Using these solutions we directly obtain 
\begin{eqnarray}
\mathcal{G}_{\xi} = \frac{\hbar \omega \xi T}{2},
\quad 
\mathcal{S}_{\xi} =
-\frac{\hbar\omega \xi T}{2} -i\hbar z_1z_2^*e^{-i\omega \xi T},\nonumber
\end{eqnarray}
and $\Lambda_\xi=i\hbar u_{\xi}v_{\xi}=i\hbar z_1z_2^*e^{-i\omega \xi T}$. The prefactor becomes
\begin{eqnarray}
\left(\frac{i}{\hbar}\frac{\partial^2 \mathcal{S}_{\xi}}
{\partial z_2^*  \partial z_1} \right)^{1/2}= e^{-\frac{i\omega \xi T}{2}}.\nonumber
\end{eqnarray}
The final result is
\begin{eqnarray}
\mathcal{K}_{\xi}(z_2^*,z_1,T) = e^{-\frac{i\omega\xi T}{2}}e^{-\frac12 |z_1|^2 -\frac12 |z_2 |^2 }e^{z_1z_2^*e^{-i\omega\xi T}},\nonumber
\end{eqnarray}
which is identical to the exact one,
\begin{eqnarray}
K_{\xi}(z_2,z_1,T) &=& \langle z_2 | e^{-i \omega\xi T 
\left( {\hat{a}^{\dag}} {\hat{a}} + \frac{1}{2} 
\right)} | z_1 \rangle 
\nonumber \\&=&
\displaystyle
e^{-\frac{i\omega\xi T}{2}}
\langle z_2 | e^{-i \omega \xi T} z_1 \rangle .
\label{OHSexato}\nonumber
\end{eqnarray}

\section{Semiclassical measure of entanglement for pure bipartite systems \label{semiP}}

We now focus on the main task of this paper, namely the derivation of a semiclassical measure of entanglement for pure bipartite systems via coherent-state propagators. In order to do so we need to extend the results of the previous section to bipartite systems. The procedure is well known~\cite{ribeiro04,garg} and its generalization for the complex conjugate of the propagator is straightforward.

We consider the coherent-state basis given by $|\mathbf z \rangle = |z_x,z_y\rangle=|z_x\rangle \otimes |z_y\rangle$, and the classical variables $\mathbf u_{\xi}= (u_x^{\xi},u_y^{\xi})$ and $\mathbf v_{\xi} = (v_x^{\xi},v_y^{\xi})$. While $\mathcal{S}_{\xi}$ has a straightforward extension, the function $\mathcal{G}_{\xi}$ requires the change
\begin{eqnarray}
\frac{\partial^2 H(v_{\xi},u_{\xi})}{\partial v_{\xi} \partial u_{\xi}} \rightarrow 
\left(
\frac{\partial^2 H(\mathbf v_{\xi},\mathbf u_{\xi})}
{\partial v_x^{\xi} \partial u_x^{\xi}} + 
\frac{\partial^2 H(\mathbf v_{\xi},\mathbf u_{\xi})}
{\partial v_y^{\xi} \partial u_y^{\xi}} \right).\nonumber
\end{eqnarray}
As far as the prefactor is concerned we need to replace the function $(i/\hbar)(\partial^2 \mathcal{S}_{\xi}/\partial z_2^*\partial z_1)$ by
\begin{eqnarray}
\det 
\left[
\frac{i}{\hbar} \left( 
\begin{array}{cc}
\frac{\partial^2 \mathcal{S}_{\xi}}{\partial z_{2x}^* \partial z_{1x}}  & 
\frac{\partial^2 \mathcal{S}_{\xi}}{\partial z_{2x}^* \partial z_{1y}} \\\\
\frac{\partial^2 \mathcal{S}_{\xi}}{\partial z_{2y}^* \partial z_{1x}}  &
\frac{\partial^2 \mathcal{S}_{\xi}}{\partial z_{2y}^* \partial z_{1y}} 
\end{array} \right)\right] \equiv
\det 
\left(\frac{i}{\hbar}
\mathbf S_{z_2^*z_1}^{\xi}\right), \quad
\label{1to2P}
\end{eqnarray}
which can be equivalently written as (see Eq.~\eqref{prefactor})
\begin{eqnarray}
\det\left(\frac{i}{\hbar}
\mathbf S_{z_2^*z_1}^{\xi}\right)= \left\{
\begin{array}{l}
\left(\det \mathrm{M_{vv}^{+}}\right)^{-1},\,\, \textrm{for}\,\,\xi=+1,\\
\left(\det \mathrm{M_{uu}^{-}}\right)^{-1},\,\, \textrm{for}\,\,\xi=-1.
\end{array}\right.\qquad
\end{eqnarray}
Notice that $\mathrm{M_{vv}^{+}}$ and $\mathrm{M_{uu}^{-}}$ are now $2\times2$ blocks of the tangent matrix. Equations~\eqref{M+} (and also their versions for $\xi=-1$) can also be extended to the case of bipartite systems by replacing each second derivative of $\mathcal S_\xi$ by a $2\times2$ matrix analogously to that of Eq.~\eqref{1to2P}.

The entanglement of a pure bipartite system composed by the subsystems $x$ and $y$, at the time $T$, can be quantified by the {\em linear entropy} of the reduced density matrix,
\be
S_{\textrm{lin}}(\hat \rho_x)=1-P(\hat \rho_x),
\label{Slin}
\ee
where $\hat \rho_x=\mathrm{Tr}_y\hat \rho$ and $\hat \rho=|\psi(T)\rangle\langle\psi(T)|$. The purity ${P}$ of the reduced density matrix $\hat \rho_x$ is defined by
\begin{eqnarray}
P (\hat \rho_x)= \mathrm{Tr}_x \{\hat\rho_x^2\}=\mathrm{Tr}_x 
\left\{\left[ \mathrm{Tr}_y\hat\rho(T)\right]^2\right\}.
\label{purity0}
\end{eqnarray}
The information about the entanglement dynamics, encoded in the linear entropy $S_{\textrm{lin}}$, is fully contained in the purity, which hence is the object of interest in this section. As we are mainly interested on the dynamical behavior of the purity, hereafter we shall denote ${P}(\hat \rho_x)$ simply by~$ P_T$.

\subsection{Semiclassical reduced density matrix}

Assuming an initial state given by $|\mathbf z_0\rangle=|z_{0x}\rangle\otimes|z_{0y}\rangle$ and a generic time-independent Hamiltonian~$\hat H$, the matrix elements of the density operator in the coherent-state representation read
\begin{eqnarray}
\langle \mathbf z_1|\hat\rho(T)|\mathbf z_2\rangle 
&=&
\langle \mathbf z_1| e^{-i\hat HT/\hbar}|\mathbf z_0 \rangle
\langle \mathbf z_0|e^{i\hat HT/\hbar} |\mathbf z_2\rangle
\nonumber \\ &=&
K_+(\mathbf z_1 , \mathbf z_0, T)~K_-(\mathbf z_0 , \mathbf z_2,T).\nonumber
\end{eqnarray}
Their semiclassical approximations are then given by 
\begin{eqnarray}
\langle \mathbf z_1|\hat\rho(T)
|\mathbf z_2\rangle_{\mathrm{semi}}
\equiv \mathcal{K}_+(\mathbf z_1^* ,\mathbf z_0,T)~\mathcal{K}_-(\mathbf z_0^*,\mathbf z_2,T),
\end{eqnarray}
which can then be evaluated by means of complex classical trajectories $(\mathbf u_{\xi}(t_\xi),\mathbf v_{\xi}(t_\xi))$ with specific boundary conditions. While, for $\mathcal{K}_+(\mathbf z_1^* ,\mathbf z_0,T)$, the boundary conditions are $\mathbf u_+'=\mathbf z_0$ and $\mathbf v_+''=\mathbf z_1^*$, for $\mathcal{K}_-(\mathbf z_0^*,\mathbf z_2,T)$, they are $\mathbf u_-''=\mathbf z_2$ and $\mathbf v_-'=\mathbf z_0^*$. Matrix elements of $\hat\rho(T)$, therefore, can be semiclassically written as functions of pairs of (generally complex) classical trajectories [$({\mathbf u}_+,{\mathbf v}_+)$ and $({\mathbf u}_-,{\mathbf v}_-)$], connected by the fact that $\mathbf u_+'= \mathbf z_0$ and $\mathbf v_-'=\mathbf z_0^*$. 

Tracing over the subsystems $y$, we obtain the matrix elements of the reduced density matrix
\begin{equation}
\begin{array}{rcl}
\displaystyle
\langle z_{1x}|\hat\rho_x(T)| z_{2x}\rangle_{\textrm{semi}} 
&=&  \displaystyle
\int \mathcal{K}_+( (z_{1x}^*,z_y^*) , \mathbf z_0, T) \\
&\times& \displaystyle 
\mathcal{K}_-(\mathbf z_0^*, (z_{2x},z_y),T)
\frac{d^2 z_y}{\pi}.
\end{array}
\label{rhored}
\end{equation}
To calculate the integral we apply the saddle point method~\cite{aguiar01,bleistein}. The critical points $(\bar z_y^*, \bar z_y)$ satisfy the relations
\be
\begin{array}{rcl}
\displaystyle
\frac{d}{d\bar z_y^*}\left[ - |\bar z_y|^2
+\frac{i}{\hbar} 
 \mathcal{S}_+( (z_{1x}^*,\bar z_y^*),\mathbf z_0, T) 
 \right] & =&0,\\\\
\displaystyle
\frac{d}{d\bar z_y}\left[ - |\bar z_y |^2  
+\frac{i}{\hbar} 
\mathcal{S}_-(\mathbf z_0^*, (z_{2x},\bar z_y),T) 
\right] &=& 0.
\end{array}\label{saddle_point}
\ee
As usual~\cite{aguiar01} we neglect $\mathcal{G}_{\xi}$ for it is a low-order term in~$\hbar$.

According to Eqs.~\eqref{ds}, the last equations imply that the critical pair of trajectories [$(\bar {\mathbf u}_+,\bar{\mathbf v}_+)$ and $(\bar{\mathbf u}_-,\bar{\mathbf v}_-)$] contributing to Eq.~\eqref{rhored} should obey the additional boundary conditions $\bar u_y^{+}(T)=\bar z_y$ and $\bar v_y^{-}(T) = \bar z_y^*$. Then, given the primary boundary conditions $\bar v_y^{+}(T)=\bar z_y^*$ and $\bar u_y^{-}(T) = \bar z_y$ it follows that among all pairs of trajectories contributing to Eq.~\eqref{rhored}, the critical ones (still complex, in general) are those for which the position and momentum in the $y$ space at the final point are real, having the same value for both trajectories, namely, $\bar u_y^{\pm}(T)=\bar z_y$ and $\bar v_y^{\pm}(T) = \bar z_y^*$. 

Expanding the integrand up to second order around the critical pair of trajectories, we get
\begin{equation}
\begin{array}{lll}
\langle z_{1x}|\hat{\rho}_x(T)| z_{2x}\rangle_{\rm semi} 
&=& 
\displaystyle
\sum_{\rm pairs}
\frac{\bar{\mathcal N} ~e^{\frac{i}{\hbar}
(\bar{\mathcal{S}}_+ + \bar{\mathcal{G}}_+ + \bar{\mathcal{S}}_- + \bar{\mathcal{G}}_-)}}
{\sqrt{\det\bar{\mathrm M}_{\rm vv}^{+}}
\sqrt{\det\bar{\mathrm M}_{\rm uu}^{-}}}~I,
\end{array}
\label{redmat}
\end{equation}
where the bar over the functions indicates that they should be calculated with the critical pairs and $\bar{\mathcal N} = e^{-|\mathbf z_0|^2-\frac12|z_{1x}|^2-\frac12|z_{2x}|^2 -|\bar{z}_y|^2}$. In addition,
\begin{equation}
I=
\int \frac{dz_y^*dz_y}{2\pi i} 
\exp\left\{ \frac12
\delta \mathrm z^T_y~
\mathrm Y~\delta \mathrm z_y\right\},\nonumber
\end{equation}
where
\begin{equation}
\delta \mathrm z^T_y = \left([z_y-\bar z_y]
\quad[z_y^* -\bar z_y^*]\right)\nonumber
\end{equation}
is the transpose of $\delta \mathrm z_y$, and
\begin{equation}
\mathrm Y=
\left(\begin{array}{cc}
\frac{i}{\hbar}\frac{\partial^2 \bar{\mathcal{S}}_-}{ \partial \bar z_y^2 } 
& -1 \\
-1 & 
\frac{i}{\hbar}\frac{\partial^2 \bar{\mathcal{S}}_+}
{\partial \{\bar{z}_y^*\}^2 } 
\end{array}\right).\nonumber
\end{equation}

The result for the Gaussian integral,
\begin{equation}
I=
\left[
1-
\left(\frac{i}{\hbar}
\frac{\partial^2 \bar{\mathcal{S}}_-}{ \partial \bar z_y^2 }\right)
\left(\frac{i}{\hbar}
\frac{\partial^2 \bar{\mathcal{S}}_+}{\partial \{\bar{z}_y^*\}^2 }\right)\right]^{-\frac12},\nonumber
\end{equation}
can be alternatively written in terms of the tangent matrix,
\be
\begin{array}{lll}
\displaystyle
\frac{i}{\hbar}\frac{\partial^2 \bar{\mathcal{S}}_-}{ \partial \bar z_y^2 }
&=&
\mathrm{h}^T_y 
\mathrm{\bar M^{-}_{vu}\left(\bar M_{uu}^{-}\right)^{-1}} 
\mathrm{h}_y,
\\\\
\displaystyle
\frac{i}{\hbar}\frac{\partial^2 \bar{\mathcal{S}}_+}
{\partial \{\bar{z}_y^*\}^2 }
&=& \mathrm{h}^T_y 
\mathrm{\bar M^{+}_{uv}\left(\bar M^{+}_{vv}\right)^{-1}}
\mathrm{h}_y,
\end{array}\nonumber
\ee
where we have defined the column matrix $\mathrm{h}_y$, whose transpose reads $\mathrm{h}_y^T=(\,0\,\,\,1\,)$. Using these expressions we get
\be
I=
\left[1- 
\mathrm{h}^T_y 
\mathrm{\bar M^{-}_{vu}\left(\bar M_{uu}^{-}\right)^{-1}} 
\mathrm{h}_y\mathrm{h}^T_y 
\mathrm{\bar M^{+}_{uv}\left(\bar M^{+}_{vv}\right)^{-1}}
\mathrm{h}_y\right]^{-\frac12}.\nonumber \\
\label{I0}
\ee

\subsection{Semiclassical Purity}

Now we proceed with the derivation of the semiclassical formula for the purity $ P_T$. For convenience, we introduce the notation 
\be
\langle z_{1x}|\hat\rho_x(T)| z_{2x}\rangle_{\mathrm{semi}}
&\equiv&
R(\mathbf v_+ '', \mathbf u_+', \mathbf v_-', \mathbf u_-'',T),
\label{R}
\ee
where we recall that contributing pairs of trajectories $(\mathbf u_{\pm}(t_{\pm}),\mathbf v_{\pm}(t_{\pm}))$ have boundary conditions $\mathbf u_+'=\mathbf z_0$, $\mathbf v_-'=\mathbf z_0^*$, $\mathbf v_+''=(z_{1x}^*, \bar z_y^*)$ and $\mathbf u_-''=(z_{2x},\bar z_y)$, and also $ u_y^{+}(T)=\bar z_y$ and $v_y^{-}(T) = \bar z_y^*$. The latter conditions state that $y$-position and $y$-momentum at time $T$ must be real, with these two classical quantities defining $\bar z_y$. For the sake of clarity, we have eliminated the bar over the trajectories involved in Eq.~\eqref{R}.

Noticing that the purity~(\ref{purity0}) can be written as
\be
 P_T =
\int \frac{d^2w_x~d^2z_x}{\pi^2}
\langle w_x|\hat\rho_x(T)| z_x\rangle
\langle z_x|\hat\rho_x(T)| w_x\rangle, \nonumber
\ee
we write the semiclassical purity as
\be
{\mathcal P}_T &=& 
\displaystyle
\int 
R(\mathbf v_+'', \mathbf u_+', \mathbf v_-', 
\mathbf u_-'',T) \nonumber
\\&\times&
\displaystyle
R(\mathbf V_+'', \mathbf U_+', 
\mathbf V_-', \mathbf U_-'',T)
\frac{d^2w_x~d^2z_x}{\pi^2},
\label{l2}
\ee
where the two contributing pairs of trajectories, $(\mathbf u_\pm,\mathbf v_\pm)$ and $(\mathbf U_\pm,\mathbf V_\pm)$, satisfy, respectively: 
\begin{equation}
\begin{array}{ll}
\mathrm{(i)}& \mathbf u_+'=\mathbf z_0, \mathbf v_-'=\mathbf z_0^*, \mathbf v_+''=(w_{x}^*, \bar z_y^*), \mathbf u_-''=(z_{x},\bar z_y),\\
& u_y^{+}(T)=\bar z_y~\mathrm{and}~v_y^{-}(T) = \bar z_y^*; \\\\
\mathrm{(ii)}& \mathbf U_+'=\mathbf z_0, \mathbf V_-'=\mathbf z_0^*, \mathbf V_+''=(z_{x}^*, \bar w_y^*), \mathbf U_-''=(w_{x},\bar w_y),\\ 
& U_y^{+}(T)=\bar w_y~\mathrm{and}~V_y^{-}(T) = \bar w_y^*.
\end{array}
\nonumber
\end{equation}
In order to find the critical trajectories [$(\bar{\mathbf u}_\pm,\bar{\mathbf v}_\pm)$ and $(\bar{\mathbf U}_\pm,\bar{\mathbf V}_\pm)$] of Eq.~(\ref{l2}), we look for its saddle points, ($ \bar w_x,\bar w_x^*$) and ($ \bar z_x,\bar z_x^*$). We find the following additional conditions:
\begin{equation}
\begin{array}{lll}
\bar V_x^{-}(T)=\bar w_x^*,& \quad &
\bar u_x^{+}(T)= \bar w_x, \\ 
\bar v_x^{-}(T)= \bar z_x^*,&\quad &
\bar U_x^{+}(T)= \bar z_x .
\end{array}
\label{adbc}
\end{equation}
Therefore, all boundary conditions that must be satisfied by the critical set of four classical trajectories contributing to $\mathcal P_T$ can be summarized as follows,
\begin{equation}
\begin{array}{lll}
\bar{\mathbf u}_+'=\mathbf z_0,&\quad \bar{\mathbf v}_+''=(\bar w_{x}^*, \bar z_y^*),&
\quad \bar{\mathbf u}_+''=(\bar w_{x}, \bar z_y),\\
\bar{\mathbf v}_-'=\mathbf z_0^*,&\quad \bar{\mathbf u}_-''=(\bar z_{x},\bar z_y),&
\quad \bar{\mathbf v}_-''=(\bar z_{x}^*, \bar z_y^*),\\
\bar{\mathbf U}_+'=\mathbf z_0,&\quad \bar{\mathbf V}_+''=(\bar z_{x}^*, \bar w_y^*),&
\quad \bar{\mathbf U}_+''=(\bar z_x, \bar w_y),\\
\bar{\mathbf V}_-'=\mathbf z_0^*,&\quad \bar{\mathbf U}_-''=(\bar w_{x},\bar w_y),&
\quad \bar{\mathbf V}_-'' = (\bar w_x^*,\bar w_y^*).
\end{array}
\label{setcond}
\end{equation}

As discussed previously, the final point of the trajectory $(\bar{\mathbf u}_+,\bar{\mathbf v}_+)$ is connected to the final point of $(\bar{\mathbf u}_-,\bar{\mathbf v}_-)$, implying the position and momentum in the $y$ direction to be real and the same for both trajectories. An equivalent  conclusion applies to $(\bar{\mathbf U}_+,\bar{\mathbf V}_+)$ and $(\bar{\mathbf U}_-,\bar{\mathbf V}_-)$. Analogously, in the $x$ direction, we see by Eqs.~\eqref{adbc} that the trajectory $(\bar{\mathbf u}_+,\bar{\mathbf v}_+)$ is connected to $(\bar{\mathbf U}_-,\bar{\mathbf V}_-)$, while $(\bar{\mathbf u}_-,\bar{\mathbf v}_-)$ is connected to $(\bar{\mathbf U}_+,\bar{\mathbf V}_+)$. This means that the trajectories contributing to $\mathcal P_T$ constitute a set of four trajectories whose final and initial conditions are mutually connected according to Eq.~(\ref{setcond}). 

A close look at these boundary conditions reveals that there exists at least one trivial set of classical trajectories satisfying all of them. It corresponds to the trajectory starting from $\tilde{\mathbf u}_+'=\tilde{ \mathbf u}_-'=\tilde{\mathbf U}_+'=\tilde{\mathbf U}_-'=\mathbf z_0$ and $\tilde{\mathbf v}_+'=\tilde{\mathbf v}_-'=\tilde{\mathbf V}_+'=\tilde{\mathbf V}_-'=\mathbf z_0^*$. Hereafter we use tilde to refer to this set of four identical trajectories which obviously satisfy, in addition, the conditions $\bar z_x=\bar w_x$ and $\bar z_y=\bar w_y$. 

Now a rather important point concerning the contributing trajectories should be identified. Consider the class of time-independent classical Hamiltonians $H(\mathbf v,\mathbf u)$ deriving from hermitian Hamiltonian operators $\hat H(\hat{\mathbf{q}},\hat{\mathbf{p}})$. In these systems a trajectory whose phase-space variables are all real at a given instant of time remains real for all times~\cite{comment3}. Since the boundary conditions given by Eqs.~\eqref{setcond} assure that the final point is real, the critical set of trajectories contributing to $\mathcal P_T$ has exclusively real trajectories. Therefore, once the initial point is completely determined there is no other solution to Eqs.~\eqref{setcond} but the trivial set discussed above.

Applying the saddle point method to expand Eq.~(\ref{l2}) around the set of real trajectories we obtain
\be
\mathcal P_T =
\mathcal I ~
R(\tilde{\mathbf v}_+'', \tilde{\mathbf u}_+', 
\tilde{\mathbf v}_-', \tilde{\mathbf u}_-'',T)
~
R(\tilde{\mathbf V}_+'', \tilde{\mathbf U}_+', 
\tilde{\mathbf V}_-', \tilde{\mathbf U}_-'',T),
\nonumber \\ \label{Ps_sets}
\ee
where
\begin{equation}
\mathcal I = \int 
\frac{d^2w_x~d^2z_x}{(2\pi i)^2}
~\exp\left\{\frac12 \delta \mathrm w^T ~\mathrm A~ \delta \mathrm w\right\},
\nonumber
\end{equation}
with
\be
\delta \mathrm w^T =
\left(\begin{array}{llll}
[w_x-\bar w_x]&
[w_x^*-\bar w_x^*]&
[z_x-\bar z_x]&
[z^*_x-\bar z^*_x]
\end{array}\right).\nonumber
\ee
The $4\times4$ matrix $\mathrm A$ can be written as
\begin{equation}
\mathrm A=
\left(
\begin{array}{cccc}
A_a+C_a&-1&0&C_c\\-1&A_{b}+C_b&C_c&0\\
0&C_c&A_a+C_a&-1\\C_c&0&-1&A_{b}+C_b
\end{array} \right),
\nonumber
\end{equation}
where
\begin{equation}
\begin{array}{l}
\displaystyle
A_{a}=\frac{i}{\hbar}
\frac{\partial^2\tilde{\mathcal{S}}_-}{\partial \{ \tilde u''_x\}^2},
\qquad
A_b=\frac{i}{\hbar}
\frac{\partial^2\tilde{\mathcal{S}}_+}{\partial \{\tilde v''_x\}^2},
\\
\displaystyle
C_a=
\left(
\frac{i}{\hbar}
\frac{\partial^2\tilde{\mathcal{S}}_-}
{\partial \tilde u''_x\partial \tilde u''_y}\right)^2
\left(
\frac{i}{\hbar}
\frac{\partial^2\tilde{\mathcal{S}}_+}{\partial \{\tilde v''_y\}^2}
\right)D,
\\
\displaystyle
C_b=
\left(
\frac{i}{\hbar}\frac{\partial^2\tilde{\mathcal{S}}_+}
{\partial \tilde v''_x\partial \tilde v''_y}\right)^2
\left(
\frac{i}{\hbar}
\frac{\partial^2\tilde{\mathcal{S}}_-}{\partial \{\tilde u''_y\}^2}
\right)D,
\\
\displaystyle
C_c =
\left(\frac{i}{\hbar}
\frac{\partial^2\tilde{\mathcal{S}}_+}
{\partial \tilde v''_x\partial \tilde v''_y}\right)
\left(\frac{i}{\hbar}
\frac{\partial^2\tilde{\mathcal{S}}_-}
{\partial \tilde u''_x\partial \tilde u''_y} \right)D,
\end{array}
\nonumber
\end{equation}
and $1/D=1- \left(\frac{i}{\hbar} \frac{\partial^2\tilde{\mathcal{S}}_-}{\partial \{\tilde u''_y\}^2}\right)\left(\frac{i}{\hbar} \frac{\partial^2\tilde{\mathcal{S}}_+}{\partial \{\tilde v''_y\}^2}\right)$. The Gaussian integral $\mathcal{I}$ then results
\begin{equation}
\begin{array}{rcl}
\mathcal{I} &=&
\displaystyle
\Big\{
\left[1-(A_a+C_a)(A_b+C_b)\right]^2
\\&-&2C_c^2
\left[1+(A_a+C_a)(A_b+C_b)\right]+C_c^4
\Big\}^{-\frac12}.
\end{array}
\label{If}
\end{equation}
Since the four trajectories are identical, we define
\be
\begin{array}{ll}
\tilde{\mathrm M}_{\rm uu} =
\tilde{\mathrm M}_{\rm uu}^{\pm}=\tilde{\mathrm M}_{\rm UU}^{\pm},
& \qquad \tilde{\mathrm M}_{\rm uv} =
\tilde{\mathrm M}_{\rm uv}^{\pm}=\tilde{\mathrm M}_{\rm UV}^{\pm},\\ \\
\tilde{\mathrm M}_{\rm vu} =
\tilde{\mathrm M}_{\rm vu}^{\pm}=\tilde{\mathrm M}_{\rm VU}^{\pm},
& \qquad \tilde{\mathrm M}_{\rm vv} =
\tilde{\mathrm M}_{\rm vv}^{\pm}=\tilde{\mathrm M}_{\rm VV}^{\pm}.
\end{array}\qquad \label{M_tilde}
\ee
Then, for $r$ or $s$ assuming $x$ or $y$,
\begin{equation}
\begin{array}{rcl}
\displaystyle
\frac{i}{\hbar}
\frac{\partial^2 \tilde{\mathcal{S}}_-}{\partial \tilde u''_r\partial \tilde u''_s }
&=& \mathrm{h}^T_r ~\mathrm{\tilde M_{vu}\tilde M_{uu}^{-1}} 
 ~\mathrm{h}_s ,\\ \\
\displaystyle
\frac{i}{\hbar}
\frac{\partial^2 \tilde{\mathcal{S}}_+}{\partial\tilde v''_r\partial \tilde v''_s } 
&=& \mathrm{h}^T_r ~
\mathrm{\tilde M_{uv}\tilde M^{-1}_{vv}}
 ~\mathrm{h}_s ,
\end{array}
\end{equation}
with $\mathrm{h}_x^T=(\,1\,\,\,\, 0\,)$. In addition, $R(\tilde{\mathbf v}_+'', \tilde{\mathbf u}_+', \tilde{\mathbf v}_-', \tilde{\mathbf u}_-'',T)=
R(\tilde{\mathbf V}_+'', \tilde{\mathbf U}_+', \tilde{\mathbf V}_-', \tilde{\mathbf U}_-'',T) \equiv \tilde R$, with
\begin{equation}
\begin{array}{rcl}
\tilde R &=& \displaystyle
\left[\det\tilde{\mathrm M}_{\rm uu}\right]^{-\frac12}
\left[\det\tilde{\mathrm M}_{\rm vv}\right]^{-\frac12}
\\
&\times&\displaystyle
\left[1- 
\mathrm{h}^T_y 
\mathrm{\tilde M_{vu}\left(\tilde M_{uu}\right)^{-1}} 
\mathrm{h}_y\mathrm{h}^T_y 
\mathrm{\tilde M_{uv}\left(\tilde M_{vv}\right)^{-1}}
\mathrm{h}_y\right]^{-\frac12}. \qquad
\end{array}
\end{equation}
Inserting the last results in Eq.~(\ref{Ps_sets}), we obtain
\be
{\mathcal P}_T
&=&
\tilde{\mathcal E}^{-1/2}\det\tilde{\mathrm M}_{\rm uu}
\det\tilde{\mathrm M}_{\rm vv}
\label{puritytrivial}
\ee
where
\be
\begin{array}{lll}
\tilde{\mathcal E}&=&
\tilde{\mathcal E}' + 
\Big[
\tilde{\mathcal E}''+\left(\det\tilde{\mathrm M}_{\rm uu}
\det\tilde{\mathrm M}_{\rm vv}-
\det\tilde{\mathrm A}
\det\tilde{\mathrm B}\right) \nonumber\\
&\times&
\left(\det\tilde{\mathrm M}_{\rm uu}
\det\tilde{\mathrm M}_{\rm vv}-
\det\tilde{\mathrm C}
\det\tilde{\mathrm D}\right) - \tilde{\mathcal E}''
\Big]^{2}\\
\tilde{\mathcal E}'&=&
-4 \left(\det\tilde{\mathrm M}_{\rm uu}
\det\tilde{\mathrm M}_{\rm vv}\det\tilde{\mathrm A}'
\det\tilde{\mathrm B}'\right)^2\\
\tilde{\mathcal E}''&=&
\left(
\det\tilde{\mathrm A}'\right)^2
\det\tilde{\mathrm B}
\det\tilde{\mathrm D}-\left(
\det\tilde{\mathrm A}'
\det\tilde{\mathrm B}'\right)^2\\&+&
\left(
\det\tilde{\mathrm B}'\right)^2
\det\tilde{\mathrm A}
\det\tilde{\mathrm C},
\end{array},\nonumber
\ee
with
\begin{equation}
\begin{array}{lll}
\left(\begin{array}{cc}
\tilde{\mathrm A}&\tilde{\mathrm D}\\
\tilde{\mathrm C}&\tilde{\mathrm B}
\end{array}\right) &\equiv&
\left(\begin{array}{cccc}
1&0&0&0\\
0&0&0&1\\
0&0&1&0\\
0&1&0&0
\end{array}\right)
\tilde{\mathrm M},\\\\
\left(\begin{array}{cc}
\tilde{\mathrm A}'&\tilde{\mathrm D}'\\
\tilde{\mathrm C}'&\tilde{\mathrm B}'
\end{array}\right) &\equiv&
\left(\begin{array}{cccc}
1&0&0&0\\
0&0&1&0\\
0&1&0&0\\
0&0&0&1
\end{array}\right)
\tilde{\mathbf M} .
\end{array}\label{ABCD}
\end{equation}

Equation~\eqref{puritytrivial} defines the general recipe for the calculation of the semiclassical purity and constitutes, therefore, the second important contribution of this paper. A crucial information emerges from this result, namely, that the semiclassical purity strongly depends on the determinant of sub-blocks of the tangent matrix. This implies the purity to be essentially determined by the stability of the (real) classical trajectories underlying the corresponding classical system. In other words, the semiclassical purity is sensitive to whether the trajectory is chaotic or regular.

Note that Eq.~\eqref{puritytrivial} results 1 for the case of noninteracting subsystems, in agreement with the result predicted by quantum theory. In this case the elements $\tilde{M}_{u_ru_s}$, $\tilde{M}_{u_rv_s}$, $\tilde{M}_{v_ru_s}$, and $\tilde{M}_{v_rv_s}$, where both $r$ and $s$ may assume $x$ and $y$, with $r\neq s$, vanish because the subspaces do not couple. Then a straightforward manipulation of Eq.~\eqref{puritytrivial} leads to the expected result.

Therefore, given the classical Hamiltonian $H(\mathbf v, \mathbf u)$ and the center $\mathbf z_0$ of the initial state, the calculation of the purity with Eq.~\eqref{puritytrivial} becomes a problem of classical mechanics. One may wonder whether the semiclassical formula is able to describe the dependence of the purity on the characteristics of the initial state other than its centroid. However, by examining Eq.~\eqref{H_xi} we realize that $H(\mathbf v, \mathbf u)$ itself has information not only about the physical interaction but also contains quantities that characterize $|\mathbf z_0\rangle$, namely its variances $b_{x,y}$ and $c_{x,y}$. 

\section{Case study: nonlinearly coupled oscillators \label{example}}

As an example of application of the formalism we show now that, using Eq.~(\ref{puritytrivial}), the short-time behavior of the purity is suitably reproduced. 

Consider the following Hamiltonian
\begin{equation}
\hat H = \hat H_x \otimes \mathbf{1}_y+ \mathbf{1}_x\otimes\hat H_y + \lambda \hat H_x \otimes \hat H_y,
\label{H_biq}
\end{equation}
where
\begin{equation}
\hat H_{r} = \frac{\hat p_{r}^2}{2m_r} +  \frac{m_r\omega_{r}^2\hat q_{r}^2}{2},\nonumber
\end{equation}
for $r=x$ or $y$. The initial state $|\psi_0\rangle=|z_{0x}\rangle\otimes|z_{0y}\rangle$ is chosen such that $|z_{0r}\rangle$ is the coherent state associated to $\hat H_{r}$. The annihilation operator $\hat{a}_r$ and its eigenvalue $z_{0r}$ are
\begin{equation}
\hat a_{r} = \frac{1}{\sqrt2}
\left(
\frac{\hat q_{r}}{b_r}+
\frac{i\hat p_{r}}{c_r}
\right) \quad \textrm{and}\quad
z_{0r} = \frac{1}{\sqrt2}
\left(
\frac{q_{0r}}{b_r}+
\frac{i p_{0r}}{c_r}
\right),
\nonumber 
\end{equation}
where $b_r=\sqrt{\hbar/(m_r\omega_r)}$ and $c_r=\sqrt{m_r \hbar \omega_r}$. $(q_{0r},p_{0r})$ gives the location of the center of the wave packet in phase space. In terms of the annihilation and creation operators the Hamiltonian is written
\be
\hat H=\hbar \Omega_x \hat{a}_x^{\dag}\hat{a}_x+\hbar \Omega_y \hat{a}_y^{\dag}\hat{a}_y+\hbar \Gamma\hat{a}_x^{\dag}\hat{a}_x \,\hat{a}_y^{\dag}\hat{a}_y+\epsilon_0,\label{Haa}
\ee
where $\Omega_r=\omega_r+\Gamma/2$, $\Gamma=\lambda\hbar\omega_x\omega_y$, and $\epsilon_0=\hbar(\omega_x+\omega_y)/2$. According to~\eqref{H_xi} the underlying classical Hamiltonian is
\begin{equation}
H(\mathbf{v},\mathbf{u}) = \hbar\Omega_x v_xu_x+\hbar\Omega_y v_yu_y+\hbar\Gamma v_xu_x\,v_yu_y+\epsilon_0.\nonumber
\end{equation}
The classical trajectories can be readily integrated and are given by
\begin{equation}
\left(\begin{array}{l}
u_x(t)\\u_y(t)\\v_x(t)\\v_y(t)
\end{array}\right)=
\left(\begin{array}{l}
u_x'e^{-\lambda_x t}\\u_y'e^{-\lambda_y t}\\
v_x'e^{+\lambda_x t}\\v_y'e^{+\lambda_y t}
\end{array}\right),
\end{equation}
where $\lambda_x = i\left(\Omega_x+\Gamma u_y'v_y' \right)$ and $\lambda_y = i\left(\Omega_y+\Gamma u_x'v_x' \right)$. The tangent matrix can be written as the product of two matrices $\mathrm{M}_1$ and $\mathrm{M}_2$ such that
\begin{equation}
\left(\begin{array}{l}
\delta u_x''\\ \delta u_y''\\
\delta v_x''\\ \delta v_y''
\end{array}\right)
=
\mathrm M_2 \mathrm M_1
\left(\begin{array}{l}
\delta u_x'\\ \delta u_y'\\ \delta v_x'\\
\delta v_y'
\end{array}\right),
\end{equation}
where
\begin{equation}
\mathrm M_1 =
\left(\begin{array}{cccc}
1&-au_x'v_y'&0&-au_x'u_y'\\
-au_y'v_x'&1&-au_y'u_x' &0\\
0&av_x'v_y'&1&av_x'u_y'\\
av_y'v_x'&0&av_y'u_x'&1
\end{array}\right),\nonumber
\end{equation}
with $a=i\Gamma T$, and
\begin{equation}
\mathrm M_2 = 
\left(\begin{array}{cccc}
e^{-\lambda_xT}&0&0&0\\
0&e^{-\lambda_yT}&0&0\\
0&0&e^{+\lambda_xT}&0\\
0&0&0&e^{+\lambda_yT}
\end{array}\right).\nonumber
\end{equation}
As we are interested just in the trajectory starting from ${\mathbf u}' = \mathbf z_0$ and ${\mathbf v}' = \mathbf z_0^*$, we obtain
\begin{equation}
\begin{array}{lll}
\det\tilde{\mathrm M}_{\rm uu} &=&
e^{-(\lambda_x+\lambda_y)T}
\left(1-a^2|z_{0x}|^2|z_{0y}|^2 \right),\\
\det\tilde{\mathrm M}_{\rm vv} &=&
e^{+(\lambda_x+\lambda_y)T}
\left(1-a^2|z_{0x}|^2|z_{0y}|^2 \right),
\end{array}\nonumber
\end{equation}
and, according to Eq.~(\ref{ABCD}),
\begin{equation}
\begin{array}{rcl}
\det \tilde{\mathrm A} &=&
a^2|z_{0x}|^2(z_{0y}^*)^2e^{-(\lambda_x-\lambda_y)T},\\
\det \tilde{\mathrm B} &=&
a^2|z_{0x}|^2(z_{0y})^2e^{+(\lambda_x-\lambda_y)T},\\
\det \tilde{\mathrm C} &=&
a^2(z_{0x}^*)^2|z_{0y}|^2e^{+(\lambda_x-\lambda_y)T},\\
\det \tilde{\mathrm D} &=&
a^2(z_{0x})^2|z_{0y}|^2e^{-(\lambda_x-\lambda_y)T},\\
\det \tilde{\mathrm A}' &=& a z_{0x}^*z_{0y}^*,\\
\det \tilde{\mathrm B}' &=& -a z_{0x}z_{0y}.
\end{array}\nonumber
\end{equation}
Then, Eq.~(\ref{puritytrivial}) becomes
\begin{equation}
{\mathcal P}_T=\frac{1+x}{\sqrt{1+6x+x^2(3+2x)^2}},
\end{equation}
where $x=|z_{0x}|^2|z_{0y}|^2\Gamma^2T^2$. It is important to compare the semiclassical result with the exact one. Using common techniques of the quantum formalism we obtain
\begin{equation}
P_T=e^{-2|z_{0x}|^2}\sum\limits_{n,m}\frac{|z_{0x}|^{2(n+m)}}{n!m!}~e^{-4|z_{0y}|^2\sin^2\left[\frac{\Gamma T(n-m)}{2}\right]}.
\end{equation}
This result is clearly different from the semiclassical one. In particular, we see that, at the instant $2\pi/\Gamma$, the quantum result predicts the total recoherence of the subsystem, i.e., $P_{2\pi/\Gamma}=1$. The semiclassical formula, however, results in a monotonically decreasing function of $T$, and as such is not able to reproduce the recoherence. On the other hand, let us consider the short-time scale ($\Gamma T \ll 1$). In this case the quantum result simplifies to
\be
P_T\simeq 1-2 \,|z_{0x}|^2\,|z_{0y}|^2\,\Gamma^2\, T^2,
\ee
which accurately agrees with the semiclassical purity ${\mathcal{P}}_T$ in this limit. 

This example revealed the limitations of our semiclassical formula. We see that the approach is not able to capture the physics of recoherences, which is associated with important quantum phenomena such as interferences and revivals. Actually, we have seen that the semiclassical purity reproduces accurately the exact result only within a very short time scale ($\Gamma T\ll 1$) that is much shorter than the one in which recoherence occurs ($\Gamma T=2\pi$). It has been shown in the literature that interference phenomena can be reproduced by semiclassical approaches involving more than one trajectory (see, for instance, Ref.~\cite{marcel}). We expect a similar strategy to be able to improve our results for longer times. This, however, requires one to carefully revisit the formalism looking for further contributing trajectories, which do not obey exactly the boundary conditions~\eqref{setcond}.

As far as the entanglement is concerned --- here measured by the linear entropy $S_{\rm lin}=1-P_T$ --- we may write
\be
S_{\rm lin}\simeq 2\,H_{int}\,\mathbb{T}_x\,\mathbb{T}_y,
\label{S_biq}
\ee
where we have defined the dimensionless time $\mathbb{T}_r\equiv~\omega_r T$. In this expression $H_{int}$ corresponds precisely to the classical version of the interaction Hamiltonian given in Eq.~\eqref{H_biq}. Notice that the short-time entanglement grows proportionally to the magnitude of the interaction, as expected. Surprisingly, however, it does not depend on $\hbar$ at all, thus corroborating our claim that the onset of the entanglement dynamics can be described in terms of classical mechanisms.

It is worth noticing that our semiclassical formula does predict a dependence on $\hbar$, in general. Consider, for instance, an arbitrary classical function $H(q,p)$. The application of usual quantization rules to this function (see, e.g., Ref.~\cite{angelo03}) produces an $\hbar$-independent operator ${\mathrm H}(\hat{q},\hat{p})$. However, the classical Hamiltonian entering in our recipe is given by $\langle z|\mathrm{H}(\hat{q},\hat{p})|z\rangle=H(q,p)+\sum_{n>0}\hbar^nf_n(q,p)$, which generally depends on $\hbar$~\cite{aguiar01}. It follows that the stability matrix and the semiclassical purity will depend on $\hbar$ as well. However, in the regime of large actions and energies this dependence manifests as a perturbation to the dynamics generated by $H(q,p)$ so that our claim remains valid.

\section{Final remarks \label{final}}

We have derived a semiclassical formula for the purity of pure bipartite systems initially prepared in a product of coherent states. Since we are here concerned only with pure states, our formula turns out to be a direct semiclassical measure of entanglement. As a preliminary step towards the development of our formalism, we have derived a unified semiclassical formula which is able to approach both propagators and their complex conjugate.

Our result for the semiclassical purity is given in terms of a very compact formula~\eqref{puritytrivial} which is shown to depend only on the trajectories of an auxiliary classical system. Specifically, the short-time entanglement dynamics is proven to depend exclusively on the elements of the tangent matrix, which defines the local stability of the classical trajectories. As a consequence the initially separable wave functions get spread and then entangle according to a rate that strongly depends on whether the corresponding classical trajectory is chaotic or regular. 

Finally, in order to illustrate the theory, the formalism has been applied to the problem of two nonlinearly coupled oscillators, whose dynamics is rich in quantum effects, such as collapses and revivals. The semiclassical approximation has shown to reproduce exactly the entanglement dynamics in the short-time regime. This is consistent with the approximations underlying the method.

Our results are in qualitative consonance with those reported in Refs.~\cite{jacquod1,jacquod2} and give additional analytical support to the widely known fact that the entanglement dynamics in the regime of short times depends on the characteristics of the classical point in phase space in which the initial state has been centered (see, e.g., Ref.~\cite{kyoko}). Moreover, they emphasize the fact that the short-time entanglement is promoted essentially by classical mechanisms, which here have been identified to be the stability of underlying classical trajectories. Our findings provide, therefore, analytical support to the numerical results of Refs.~\cite{angelo04,angelo05} which show that it is possible to mimic the entanglement dynamics in terms of entropic measures defined in the Liouvillian theory.

The natural continuation of this work consists in extending the formalism to spin degrees of freedom. Moreover, even though we have assumed the initial state to be a product of coherent states the generalization of the semiclassical purity to arbitrary initial states is possible. Research on these topics are now in progress.\\

\acknowledgments
A.D.R. and R.M.A. acknowledge financial support from CNPq/Brazil. We would like to thank M. A. M. de Aguiar for a careful reading of this paper and for valuable suggestions. We also thank J. G. P. Faria, G. Q. Pellegrino, and M. V. S. Bonan\c{c}a for helpful discussions.

%


\begin{thebibliography}{99}
\bibitem{comment1} 
There is another famous example of the connection between classical concepts and exclusively quantum effects, namely, the problem of quantization rules, in which the quantum energy spectrum is built up from classical orbits~\cite{gutz}.
\bibitem{comment2} 
It is also possible to think of the change $T\to \xi T$, since $K_{\xi}(z_2,z_1,T)=K(z_2,z_1,\xi T)$. We prefer the replacement $\hat{H}\to\xi \hat{H}$ because the corresponding changes in the formalism occur in a more direct and easily justifiable way.
\end{thebibliography}
\end{document}